\documentclass[prl,twocolumn,superscriptaddress,nofootinbib,floatfix]{revtex4-2}

\usepackage{graphicx,color,dcolumn,booktabs,bm}
\usepackage{amsmath,amssymb}
\usepackage{txfonts}
\usepackage{leftindex}
\usepackage{ulem}
\usepackage{feynmf}   %{feynmp}
\usepackage{slashed}  %for Feynman symbols
\usepackage{cases}
\usepackage{color}
\usepackage{multirow}
\usepackage{threeparttable}
\usepackage{epstopdf}
\usepackage{enumerate}
\usepackage[colorlinks=true,
            citecolor=blue,
            anchorcolor=red,
            menucolor=red,
            linkcolor=red,
            filecolor=red,
            runcolor=red,
            urlcolor=blue,
            frenchlinks=red]{hyperref}
\usepackage{appendix}
\usepackage{orcidlink}

\graphicspath{{Figures/}}

\newcommand{\lzu}{\affiliation{School of Physical Science and Technology, Lanzhou University, Lanzhou 730000, China}}

\newcommand{\lctp}{\affiliation{Lanzhou Center for Theoretical Physics,
Key Laboratory of Theoretical Physics of Gansu Province,\\
Key Laboratory of Quantum Theory and Applications of MoE,\\
Gansu Provincial Research Center for Basic Disciplines of Quantum Physics, Lanzhou University, Lanzhou 730000, China}}

\newcommand{\Isotopes}{\affiliation{MoE Frontiers Science Center for Rare Isotopes, Lanzhou University, Lanzhou 730000, China}}

\newcommand{\CSR}{\affiliation{Research Center for Hadron and CSR Physics, Lanzhou University and Institute of Modern Physics of CAS, Lanzhou 730000, China}}

\begin{document}

\title{Unified coupled-channel description for the five near-threshold structures  $\psi(3770)$, $G(3900)$, $R(3760)$, $R(3780)$ and $R(3810)$ from $e^+e^-$ annihilation}

\author{Ri-Qing Qian\orcidlink{0000-0002-5352-1243}}\email{qianrq@lzu.edu.cn}
\lzu\lctp\CSR
\author{Xiang Liu\orcidlink{0000-0001-7481-4662}}\email{xiangliu@lzu.edu.cn}
\lzu\lctp\Isotopes\CSR

\begin{abstract}
   The recent observation of multiple near-threshold structures in $e^+e^-$ annihilation—including $\psi(3770)$, $G(3900)$, $R(3760)$, $R(3780)$, and $R(3810)$—reveals limitations in existing models of charmonium and exotic hadrons. In this paper, we propose a unified coupled-channel description that simultaneously incorporates all five near-threshold structures using parameters constrained by hadron spectroscopy. Our model accurately reproduces the line shapes in both $D\bar{D}$ and nonopen-charm hadron (nOCH) channels, resolves the asymmetric profile of $\psi(3770)$, and produces the $G(3900)$ bump. We identify the $R(3780)$ as the dominant manifestation of $\psi(3770)$, attribute $R(3760)$ to $D\bar{D} \to \text{nOCH}$ rescattering and suggest that $R(3810)$ arises from coupling between the $\psi(1D)$ state and the $h_c\pi\pi$ channel. The significant non-$D\bar{D}$ decay of $\psi(3770)$ is explained by its near-threshold position. This analysis provides a coherent, unquenched framework centered on a charmonium $\psi(1D)$ core for near-threshold phenomena and offers a predictive approach extendable to bottomonium systems and future data from Belle II.
\end{abstract}
%\date{\today}
\maketitle

\newcommand{\vect}[1]{\boldsymbol{#1}}

\noindent{{\it Introduction}.---}Since 2003, the discovery of numerous new hadronic states in experiments has spurred a renaissance in hadron spectroscopy (see review articles~\cite{Chen:2016qju,Guo:2017jvc,Liu:2019zoy,Brambilla:2019esw,Chen:2022asf,Wang:2025dur}). These observations are closely linked to our understanding of nonperturbative aspects of strong interactions—a central challenge in modern particle physics. Among these, charmoniumlike $XYZ$ states have attracted particular interest. A common feature of these states is their proximity to various hadronic thresholds. A notable example is the $X(3872)$~\cite{Belle:2003nnu}, often described as a “superstar” due to its location near the $D\bar{D}^*$ threshold, which has garnered widespread attention in the field.  In fact, near-threshold phenomena—exemplified by the deuteron in nuclear physics~\cite{Urey:1932gik} and Feshbach resonances in cold atoms~\cite{Chin:2010crf}—are popular in whole physics and reflect fundamental dynamical mechanisms.

In recent years, enhanced experimental precision has enabled detailed measurements of the cross sections for $e^+e^-$ annihilation into both open-charm and noncharm final states, unveiling a range of structures near open-charm thresholds. These include the well-established charmonium state $\psi(3770)$, as well as the newly reported resonances $G(3900)$, $R(3760)$, $R(3780)$, and $R(3810)$ (see Table~\ref{tab:structures}).
As the first charmonium state above the open-charm threshold, the $\psi(3770)$ displays a number of anomalous properties. Though traditionally assigned to the $1^3D_1$ $c\bar{c}$ states, the central value of the measured mass of the $\psi(3770)$ deviates from the predictions of various potential models~\cite{Eichten:1978tg,Godfrey:1985xj,Wang:2019mhs}. Also, it exhibits a significant nonopen-charm hadron (nOCH) decay fraction of approximately 15\%~\cite{BES:2007cev,BES:2008vad}, in contrast to the nearly negligible nonopen-charm hadron partial widths of the $\psi(3686)$ and $J/\psi$. Beyond its anomalous mass and decay behaviors, the $\psi(3770)$ line shape deviates from a standard Breit-Wigner distribution—a discrepancy that has led to proposal that two distinct structures might be present~\cite{BES:2008wyz}. 
In addition to the structures near $\psi(3770)$, a resonancelike feature denoted as $G(3900)$ has been observed just above the $\psi(3770)$ peak in the $e^+e^-\to D\bar{D}$ cross section~\cite{BESIII:2024ths,BaBar:2006qlj,BaBar:2008drv,Belle:2007qxm}. In response to these novel near-threshold phenomena, theoretical attempts have been made from various perspectives, such as exotic hadronic states~\cite{Lin:2024qcq} and hadron scattering theory~\cite{Cao:2014qna,Du:2016qcr,Uglov:2016orr,Husken:2024hmi,Ye:2025ywy,Nakamura:2023obk}. 
Recently, both the hadronic cross section~\cite{BESIII:2023oql} and the nonopen-charm hadron cross section~\cite{BESIII:2023bed} unveil intricate structures in the $3.7-3.82$ GeV region, which are well described by introducing three near-threshold structures: the $R(3760)$, $R(3780)$, and $R(3810)$. Clearly, this indicates a redundancy of near-threshold structures, as they cannot be comprehensively described under either the conventional charmonium or exotic hadron state frameworks.
\begin{table}[htbp]
  \caption{\label{tab:structures} 
  The structures observed in $e^+e^-$ annihilation near the $D\bar{D}$ threshold.}
  \begin{ruledtabular}
    \begin{tabular}{cccc}
        Structure & $\psi(3770)$  & $G(3900)$  & $R(3760)$, $R(3780)$ and $R(3810)$        \\ 
        \hline\
        Channel    & $D\bar{D}$    & $D\bar{D}$~\cite{BESIII:2024ths,BaBar:2006qlj,BaBar:2008drv,Belle:2007qxm} & nOCH /  hadrons~\cite{BESIII:2023oql,BESIII:2023bed}    \\
    \end{tabular}
  \end{ruledtabular}
\end{table}

Facing this challenge, we propose in this paper a unified coupled-channel description for the five near-threshold structures—$\psi(3770)$, $G(3900)$, $R(3760)$, $R(3780)$, and $R(3810)$. Our approach is based on a coupled-channel framework with most input parameters constrained by hadron spectroscopy studies, ensuring a self-consistent and robust theoretical model. Our model accurately reproduces the line shapes of the cross sections for $e^+e^-$ annihilation into both $D\bar{D}$ and nonopen-charm hadron final states. It successfully captures the asymmetric line shape of the $\psi(3770)$ and the resonancelike structure denoted as $G(3900)$. Our analysis identifies the $R(3780)$ as the $\psi(3770)$ resonance, attributes the $R(3760)$ to $D\bar{D} \to \mathrm{nOCH}$ rescattering dynamics, and suggests that the $R(3810)$ may stem from coupling between the $\psi(1D)$ state and the $h_c\pi\pi$ channel. The sizable nonopen-charm hadron branching fraction of the $\psi(3770)$ is naturally explained by its mass lying very close to the $D\bar{D}$ threshold.

Understanding these near-threshold phenomena marks a shift from a quenched to an unquenched picture in high-precision hadron spectroscopy.
Moreover, the coupled-channel framework presented here is not limited to the charmonium sector; it can be directly applied to $e^+e^-$   annihilation into open-bottom and nonopen-bottom hadrons. This extension represents a promising direction for future research, particularly with the high-statistics data expected from Belle II.

\noindent{{\it The $\psi(3770)$ and $G(3900)$ in the $D\bar{D}$ channel}.---}Our starting point is a coupled-channel framework that incorporates the bare $c\bar{c}$ core, hadron-hadron kinematics, and the transition interaction between the  bare $c\bar{c}$ states and various hadronic channels. The Hamiltonian  is expressed as 
\begin{equation}
    H = H_{c\bar{c}} + H_{c\bar{c}\leftrightarrow hadrons} + H_{hadrons}\,.
\end{equation}
The $H_{c\bar{c}}$ can be modeled using a conventional potential model~\cite{Godfrey:1985xj}, with the $c\bar{c}$ eigenstate $|\psi_n\rangle$ denoted as bare states. The hadronic Hamiltonian $H_{hadrons}$ include kinematics and interaction of hadronic channels,
\begin{equation}
    H_{hadrons} = \sum_i \left(m_i^{\mathrm{th}} + \frac{p_i^2}{2\mu_i}\right) + \sum_{i,j}V_{ij}^{direct}(\vect{p},\vect{p}') \,.
\end{equation}
When restricted to the space of hadron state $\{|i;\vect{p}\rangle\}$, the transition term $H_{c\bar{c}\leftrightarrow hadrons}$ effectively induces an $s$-channel potential between hadron states:
\begin{equation}
    V_{ij}^s(\vect{p},\vect{p}') = \sum_n \frac{f_{in}(\vect{p})f_{nj}(\vect{p}')}{E-E_n} \,,
\end{equation}
where $f_{n,i}(\vect{p}) \equiv \langle \psi_n|H_{c\bar{c}\leftrightarrow \mathrm{hadrons}}| i;\vect{p} \rangle$ denotes the transition matrix element between the $n$th bare $c\bar{c}$ state and the $i$th hadronic channel. $E_n$ represents the bare mass of the $c\bar{c}$ state, which is shifted to a complex physical mass $E-i\,\Gamma/2$ for the observed resonance after incorporating coupled-channel dynamics. The total interaction is
\begin{equation}
    V_{ij} = V_{ij}^s + V_{ij}^{direct} \,,
\end{equation}
In the following analysis, we neglect the direct hadron-hadron interaction $V_{ij}^\text{direct}$ unless explicitly required, for the following reasons. First,  $e^+e^-$ predominantly couple to vector $c\bar{c}$ states, making the $c\bar{c}$-hadron coupling ($V^s_{ij}$) essential for modeling production, while the direct hadron-hadron interaction $V_{ij}^{direct}$ only serves as a modification to the production cross section. Second, the explicit form of $V_{ij}^{direct}$ is generally more model-dependent and its effect is relatively difficult to account for.

Since the $s$-channel potential is separable, the corresponding Lippmann-Schwinger equation can be solved exactly, yielding the  $T$-matrix 
\begin{equation}
    T_{ij}(E;\vect{p},\vect{p}') = f_i^\dagger(\vect{p})\frac{\lambda}{I-\lambda\sum_{k}I_k(E)}f_j(\vect{p}') \,,
\end{equation}
where $f_i(\vect{p})=\left(f_{1i}(\vect{p}'),f_{2i}(\vect{p}'),...\right)$ and $\lambda = {\rm diag}(1/(E-E_1),1/(E-E_2),...)$. 
The term $I_k(E)$ encapsulates the contribution from the $k$-th hadron channels to the pole positions of physical states and is given by
\begin{equation}\label{eq:Ik}
    I_k(E) = \int d^3q \frac{f_k(\vect{q})f_k^\dagger(\vect{q})}{E-m_k^{\rm{th}} - q^2/(2\mu_k)} \,.
\end{equation}
We focus on the energy region just above the $D\bar{D}$ threshold. The bare states $\psi(1D)$ and $\psi(3S)$ are included, with masses fixed at 3.82 GeV and 4.10 GeV, respectively, as obtained from a potential model calculation~\cite{Godfrey:1985xj}. Although the $\psi(2S)$ lies slightly below the $D\bar{D}$ threshold, is is also included with a mass of 3.688 GeV.

Among the various hadronic channels, the open-charm channels $D^{(*)}\bar{D}^{(*)}$ play the dominant role, accounting for nearly the entire decay width of $c\bar{c}$ states above the open-charm threshold. The transition function $f_{i,n}(\vect{p})$ is calculated using the quark pair creation (QPC) model~\cite{Micu:1968mk,LeYaouanc:1972vsx}, which describes the coupling between a $c\bar{c}$ state and a $D^{(*)}\bar{D}^{(*)}$ pair via creation of a light-quark pair with the quantum numbers $J^{PC}=0^{++}$. The transition function takes the form
\begin{equation}
f_{i,n}(\vect{p}) = \gamma_n \mathcal{M}^{QPC}_{M_n\to \{M_i\}}(\vect{p})e^{-\frac{\vect{p}^2}{2\Lambda^2}} \,,
\end{equation}
where an exponential form factor is introduced to suppress high-momentum contributions inherent in the original QPC model~\cite{Morel:2002vk,Ortega:2016mms,Yang:2021tvc,Deng:2023mza,Wang:2024ytk}. The parameter $\gamma_n$ quantifies the strength of quark pair creation.
For $\psi(2S)$, we include its coupling to $D\bar{D}$, though its mass lies below the $D\bar{D}$ threshold; Couplings to other open-charm channels are neglected due to their higher threshold. For $\psi(1D)$ and $\psi(3S)$, coupling to 
$D\bar{D}^{(*)}$ and $D^{(*)}\bar{D}^{(*)}$, respectively, are included. The details of the QPC amplitudes can be found in Supplemental Material (SM)~\cite{Supplemental}.

To connect with experimental cross sections, we also introduce the coupling $f_{n,e^+e^-}$ between bare states and the $e^+e^-$ channel. The value of $f_{n,e^+e^-}$ is related to the dilepton width of the bare state within the potential model via
\begin{equation}
    \Gamma^{e^+e^-}_{n} = 8\pi^2 \frac{|\vect{p}^{cm}|}{m_n}E_e^2 \frac{1}{2J_n+1} \sum_{spin} |f_{n,e^+e^-}|^2 \,,
\end{equation}
where $E_e$ is the electron energy.
Clearly, $I_{e^+e^-}(E)$ is negligibly small compared to contribution from open-charm channels. 
The on-shell $T$-matrix element relevant to $e^+e^-\to D\bar{D}$ is 
\begin{equation}\label{eq:TDD}
    T_{D\bar{D},e^+e^-}(E+i\epsilon;\vect{p}_{D}^{cm}) = \frac{f_{D\bar{D}}(\vect{p}_D^{cm})\lambda f_{e^+e^-}}{I-\lambda (I_{D\bar{D}} + I_{D\bar{D}^*} + I_{D^*\bar{D}^*}) + i\epsilon} \,,
\end{equation}
and the cross section in our nonrelativistic convention is 
\begin{equation}
    \frac{d\sigma^{D\bar{D}}}{d\Omega} = (2\pi)^4\frac{\sqrt{s}|\vect{p}_D^{cm}|}{8}\frac{1}{4}\sum_{spin}|T_{D\bar{D},e^+e^-}|^2 \,.
\end{equation}

\begin{figure}[htbp]
  %\centering
  \includegraphics[width=8.4cm]{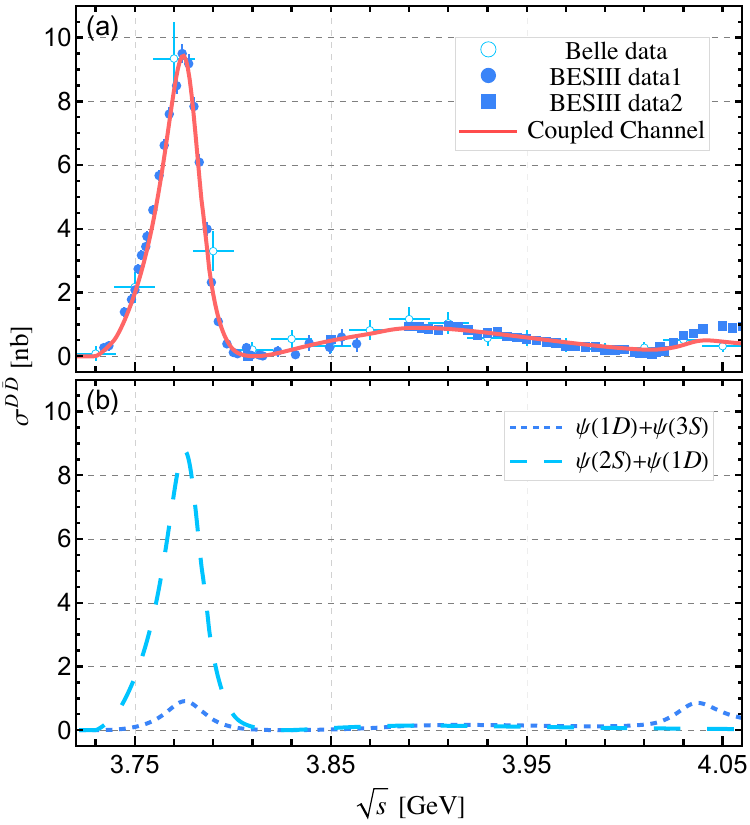}
  \caption{(a) Coupled-channel description of the $e^+e^-\to D\bar{D}$ cross section. The red solid line represents the full model including $\psi(2S)$, $\psi(1D)$, and $\psi(3S)$. Experimental data are from Belle~\cite{Belle:2007qxm} and BESIII~\cite{Husken:2024hmi,BESIII:2024ths}.
  (b) Cross section when either $\psi(2S)$ or $\psi(3S)$ is removed.}\label{fig:1D}
\end{figure}

Figure~\ref{fig:1D} shows the comparison of the coupled-channel calculation with experimental data; the corresponding parameters are listed in Table~\ref{tab:parameter}. The dilepton widths for the bare states are
\begin{equation}
    \Gamma^{e^+e^-}_{2S} = 5.50\; \mathrm{keV},\;\Gamma^{e^+e^-}_{1D} = 0.02\;\mathrm{keV},\;\Gamma^{e^+e^-}_{3S} = 4.91\;\mathrm{keV}\,.\nonumber
\end{equation}
These values are comparable to prediction from the potential model~\cite{Godfrey:1985xj}. The asymmetric line shape of the $\psi(3770)$ is naturally reproduced within this framework. Specifically, the $\psi(2S)-D\bar{D}-\psi(1D)$ coupling leads to a relatively large dilepton width for $\psi(3770)$. Meanwhile, the $\psi(1D)-D\bar{D}^{(*)}-\psi(3S)$ coupling induces an interference that produces a visible bump near 3.9 GeV in the $D\bar{D}$ channel. Removing either $\psi(2S)$ or $\psi(3S)$ eliminates this structure, indicating  a nontrivial interference pattern.

The poles corresponding to physical resonances are obtained via analytic continuation from the physical region to the relevant Riemann sheet. On the sheet connected to the physical region, we find three poles near the real axis (see Table~\ref{tab:poles}), corresponding to the $\psi(3686)$, $\psi(3770)$, and $\psi(4040)$. The inclusion of $D\bar{D}$ coupled-channel dynamics shifts the mass of the bare $\psi(1D)$ state from 3.82 GeV to that of physical $\psi(3770)$.
No additional pole is found near 3.9 GeV in the vicinity of the physical axis.

\begin{table}
  \caption{\label{tab:parameter}
  The parameters in the coupled-channel model. The coupling $f_{n,e^+e^-}$ are given in units of GeV$^{-\frac{1}{2}}$.}
  \begin{ruledtabular}
    \begin{tabular}{ccccccc}
        $\gamma_{2S}$ & $\gamma_{1D}$ & $\gamma_{3S}$ & $f_{\psi(2S),e^+e^-}$             & $f_{\psi(1D),e^+e^-}$            & $f_{\psi(3S),e^+e^-}$ & $\Lambda$ (GeV)\\
        \hline
        3.54          & 6.51           & 6.15         & $2.48\times 10^{-4}$ & $0.15\times 10^{-4}$ & $2.11\times 10^{-4}$ & 0.65  \\
    \end{tabular}
  \end{ruledtabular}
\end{table}

\begin{table}[htbp]
  \caption{\label{tab:poles} 
  Poles in the coupled-channel model. The Riemann sheets are labeled according to the sheets used for the $D\bar{D}$, $D\bar{D}$, and $D^*\bar{D}^*$ channels. For example, $(I, II, I)$ denotes the first Riemann sheet for the $D\bar{D}$ channel, the second sheet for the $D\bar{D}^*$ channel, and the first sheet for the $D^*\bar{D}^*$ channel.}
  \begin{ruledtabular}
    \begin{tabular}{cccc}
        Riemann sheet          & $(I,I,I)$    & $(II,I,I)$     & $(II,II,II)$         \\ 
        \hline
        Pole (GeV)             & 3.686        & $3.778-0.012i$ & $4.031-0.024i$    \\
        Physical state          & $\psi(3686)$ & $\psi(3770)$   & $\psi(4040)$      \\
    \end{tabular}
  \end{ruledtabular}
\end{table}

\begin{figure}[htbp]
  \centering
    \includegraphics[width = 8.4 cm]{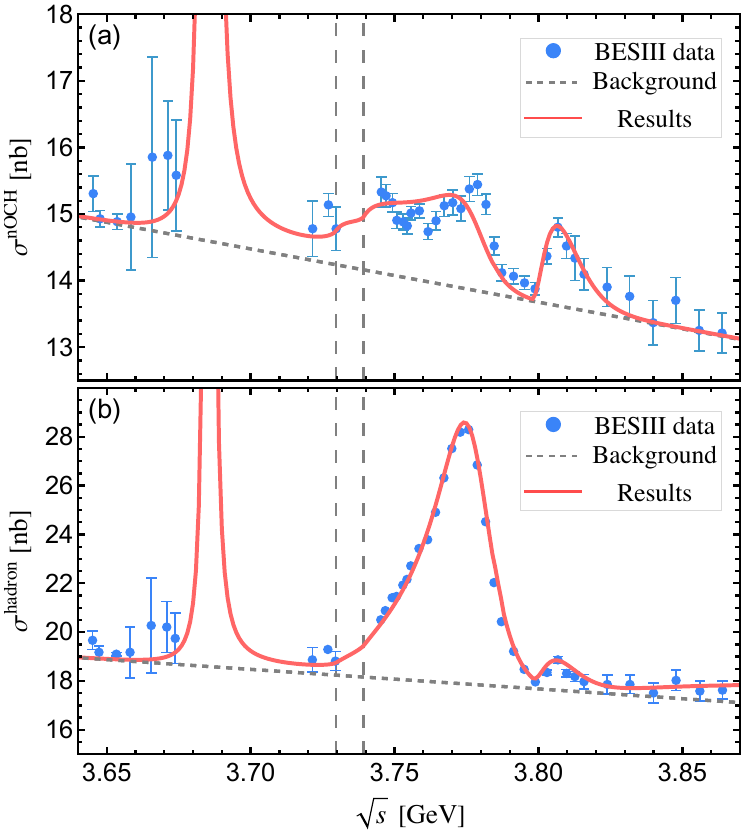}
    \caption{Coupled-channel description of (a) the $e^+e^-\to$ nonopen-charm hadron and (b) the inclusive hadronic cross sections. Vertical dashed lines indicate the $D^0\bar{D}^0$ and $D^+\bar{D}^-$  thresholds. nonopen-charm hadron data are from Ref.~\cite{BESIII:2023bed}, inclusive hadronic data are from Ref.~\cite{BESIII:2023oql}.}\label{fig:hadron}
\end{figure}

\noindent{{\it The $R(3760)$, $R(3780)$ and $R(3810)$ in the nonopen-charm hadron channel}.---}We now extend the coupled-channel framework to include nonopen-charm hadron channels. The $R(3780)$ structure, with mass and width similar to those of the conventional $\psi(3770)$, can be identified with $\psi(3770)$. 
In contrast, the $R(3760)$ structure appear near the $D\bar{D}$ threshold in the nonopen-charm hadron cross section (See Fig.~\ref{fig:hadron}) and lacks a clear $c\bar{c}$ counterpart. A natural question is whether the coupling between bare $c\bar{c}$ states and the $D\bar{D}$ channel could produce a $D\bar{D}$ threshold cusp in the nonopen-charm hadron channel when $c\bar{c}$-nOCH coupling are included. We conjecture that {\it a direct coupling between bare $c\bar{c}$ states and nonopen-charm hadron channel cannot yield any significant structure near the $D\bar{D}$ threshold.}

To clarify this point, consider the energy region very close to the $D\bar{D}$ threshold. When including $c\bar{c}$-nonopen-charm hadron coupling, the scattering amplitude can be approximated as:
\begin{eqnarray}
    T_{\text{nOCH},e^+e^-}^{0}(E+i\epsilon) \approx  \frac{f_{\text{nOCH}}\left(m_{D\bar{D}}^{\mathrm{th}}\right)\lambda f_{e^+e^-}}{I-\lambda \left(I_{D\bar{D}}(E+i\epsilon)- \sum_{k\neq D\bar{D}} I_{k}\left(m_{D\bar{D}}^{\rm{th}}\right) \right)} \,.\nonumber
\end{eqnarray}
Near the $D\bar{D}$ threshold, the only significant energy dependence comes from $I_{D\bar{D}}(E)$. The  contribution from $I_{D^{(*)}\bar{D}^*}(E)$ and $I_{\text{nOCH}}(E)$ are smooth in this region due to  their higher thresholds. The loop function $I_{D\bar{D}}$ behaves as
\begin{eqnarray}
    I_{D\bar{D}}(E+i\epsilon) &\approx& -2\mu_{D\bar{D}}\int dq (f^\dagger_{D\bar{D}} f_{D\bar{D}})(q) \nonumber\\
    & & - i\pi\mu_{D\bar{D}}k_{D\bar{D}}\, (f^\dagger_{D\bar{D}} f_{D\bar{D}})(k_{D\bar{D}}) 
\end{eqnarray}
with $k_{D\bar{D}}=\sqrt{2\mu_{D\bar{D}}E_{D\bar{D}}}$. Although the square root function could, in principle, produce a cusp, the coupling $f_{D\bar{D}}$ is in $P$-wave, leading to the behavior $f_{D\bar{D}} \approx a k_{D\bar{D}}$ as $E_{D\bar{D}}\to 0$. Consequently, the imaginary part is suppressed by $f^\dagger_{D\bar{D}}f_{D\bar{D}}\propto (2\mu_{D\bar{D}}E_{D\bar{D}})$, rendering the cusp effect negligible. This contrasts sharply with the case of an $S$-wave coupling, where $f^\dagger_if_i\approx \text{constant}$ and a pronounced cusp appears.

Since direct $c\bar{c}$-nonopen-charm hadron coupling cannot produce a noticeable threshold structure, we instead introduce a coupling between the $D\bar{D}$ and nonopen-charm hadron channels, corresponding to a rescattering process $D\bar{D}\to {\rm nOCH}$. This is modeled by
\begin{equation}
    V_{\text{nOCH},D\bar{D}}(\vect{q}) = g_{\text{nOCH}}qe^{-q/\Lambda_{\text{nOCH}}} Y_1^{1*}(\Omega_q)\,,
\end{equation} 
where $q$ is the momentum of the $D\bar{D}$ system and $Y_1^1$ is the spherical harmonic function. The momentum dependence of nonopen-charm hadron channels can be neglected due to their lower thresholds. This form captures two essential features:  (1) the $P$-wave nature of the coupling to the vector charmonium state, leading to $V_{\text{nOCH},D\bar{D}}\propto qY_1^{1*}(\Omega_q)$ as $q\to 0$; and (2) the characteristic momentum scale $\Lambda_{\text{nOCH}}$ of the rescattering.
The $T$-matrix for nonopen-charm hadron production is then
\begin{eqnarray}
    T_{\text{nOCH},e^+e^-}(E+i\epsilon) &=& V_{\text{nOCH},D\bar{D}}G^{D\bar{D}}(E+i\epsilon)T_{D\bar{D},e^+e^-}(E+i\epsilon) \nonumber\\
        &=& \int d^3q \frac{V_{\text{nOCH},D\bar{D}}(\vect{q}) T_{D\bar{D},e^+e^-}(E+i\epsilon;\vect{q})}{E-m_{D\bar{D}}^{\rm{th}}-q^2/(2\mu_{D\bar{D}})+i\epsilon} \,, \nonumber 
\end{eqnarray}
where $T_{D\bar{D},e^+e^-}$ is approximated by the amplitude in Eq.~\eqref{eq:TDD}\footnote{Strictly speaking, the $T_{D\bar{D},e^+e^-}$ should be obtained by solving the full LS equation, including the $V_{\text{nOCH},D\bar{D}}$ interaction. However, this requires knowledge of the explicit momentum dependence of nOCH channel $V_{\text{nOCH},D\bar{D}}(\vect{p}_{\text{nOCH}},\vect{p}_{D\bar{D}})$, which introduces additional parameters. As a first-order approximation, we can approximate $T_{D\bar{D},e^+e^-}$ as the $T$ matrix calculated in the absence of $V_{\text{nOCH},D\bar{D}}$. Additional discussion of the full LS equation is provided in SM~\cite{Supplemental}.}. 
To model the collective effect of multiple nonopen-charm hadron channels with thresholds well below $D\bar{D}$, we  introduce an effective two-body channel with
\begin{equation}
    \mu_{\text{nOCH}} = 0.7\; \mathrm{GeV},\quad m^{\mathrm{th}}_{\text{nOCH}} = 3.0\; \mathrm{GeV}\,.
\end{equation}

In addition to the $R(3760)$ near the $D\bar{D}$ threshold,
a small peaklike structure $R(3810)$ is observed around 3.81 GeV in the nonopen-charm hadron channel, just above the $h_c\pi\pi$ threshold and with no counterpart in $D\bar{D}$. This suggests a contribution from $e^+e^-\to h_c\pi\pi$. We therefore include an $h_c\pi\pi$ channel in the model. Lacking theoretical constraints on its coupling to $c\bar{c}$ states, and noting that the bare $\psi(1D)$ mass (3.82 GeV) is close to the $h_c\pi\pi$ threshold, we introduce a $\psi(1D)$-$h_c\pi\pi$ coupling. Assuming the two pions are in relative $P$-wave ($L_{\pi\pi}=1$), the transition matrix element as
\begin{eqnarray}
    f_{\psi(1D),h_c\pi\pi}(\vect{q},\vect{p}) &=& \frac{g_{h_c\pi\pi}}{\sqrt{4\pi}}pe^{-(q^2+p^2)/\Lambda_{h_c}^2}\frac{1}{\sqrt{2}}\left(Y_1^{0*}(\Omega_p) - Y_1^{1*}(\Omega_p) \right) \,, \nonumber
\end{eqnarray}
where $\vect{q}$ is the $h_c$ momentum in the c.m. frame, and $\vect{p}$ is the relative momentum between the pions. The contribution to the $T$-matrix is
\begin{eqnarray}
    I_{h_c\pi\pi}(E) &=& \int d^3q d^3p \frac{f_{h_c\pi\pi}(\vect{q},\vect{p})f_{h_c\pi\pi}^\dagger(\vect{q},\vect{p})}{E-m_{h_c\pi\pi}^{\rm{th}}-q^2/(2\mu_{h_c(\pi\pi)})-p^2/(2\mu_{\pi\pi})} \,,\nonumber
\end{eqnarray}
where $f_{h_c\pi\pi}=(0,f_{1D,h_c\pi\pi}(\vect{q},\vect{p}),0)$, since couplings of $\psi(2S/3S)$ to $h_c\pi\pi$ are neglected. The full $T$-matrix is approximated as 
\begin{equation}
    T_{ij} = f_i\frac{\lambda}{I-\lambda (I_{D\bar{D}} + I_{D\bar{D}^*} + I_{D^*\bar{D}^*} + I_{h_c\pi\pi} + I_{2S})}f_j \,,
\end{equation}
where a small constant term $I_{2S} = i\,\delta_{2S} {\rm diag}(1,0,0)$ is introduced to give $\psi(2S)$ a small width of $\sim 300$ keV~\cite{ParticleDataGroup:2024cfk}. Its value is determined by the relation
\begin{equation}
    \delta_{2S} = \Gamma_{2S}/2 \approx 150\;\mathrm{keV}  \nonumber\,.
\end{equation}
The $D^0\bar{D}^0$ and $D^+D^-$ channels are treated separately. The nonopen-charm hadron cross section is then
\begin{equation}
    \sigma^{\text{nOCH}} = \sigma^{D\bar{D}\to \text{nOCH}} + \sigma^{h_c\pi\pi} \,, 
\end{equation}
and the inclusive hadronic cross section is
\begin{equation}
    \sigma^{hadron} = \sigma^{D\bar{D}} + \sigma^{D\bar{D}\to \text{nOCH}} + \sigma^{h_c\pi\pi} \,.
\end{equation}

The final results for the nonopen-charm hadron cross section, including both $V_{\text{nOCH},D\bar{D}}$ and $f_{\psi(1D),h_c\pi\pi}$ couplings, are shown in Fig.~\ref{fig:hadron}. The additional parameters are
\begin{eqnarray}
    g_{\text{nOCH}}=11.5\; \mathrm{GeV}^{-1/2}\,,\quad && \Lambda_{\text{nOCH}} = 90\; \mathrm{MeV} \,, \nonumber\\
    g_{h_c\pi\pi} = 8\times 10^{3}\; \mathrm{GeV}^{-1}\,,\quad && \Lambda_{h_c} = 40\;\mathrm{MeV} \,.
\end{eqnarray}
The uncertainty of parameters $\Lambda_{\mathrm{nOCH}}$ and $\Lambda_{h_c}$ is discussed in SM~\cite{Supplemental}.The structures just above the $D\bar{D}$ threshold arise from $D\bar{D} \to \text{nOCH}$ rescattering. The scale $\Lambda_{\text{nOCH}} \approx 0.09~\mathrm{GeV}$ restricts the process to small relative momenta ($\lesssim 90~\mathrm{MeV}$), naturally explaining the significant nonopen-charm hadron decay width of the $\psi(3770)$. No direct $\psi(1D)$–nOCH coupling is required, consistent with the small non-$D\bar{D}$ width of the $\psi(3686)$. The large nonopen-charm hadron branching fraction of the $\psi(3770)$ is due to its proximity to the $D\bar{D}$ threshold, enabling decay via rescattering $D\bar{D} \to \text{nOCH}$. Similar structures may appear in $e^+e^-\to J/\psi X$ cross section~\cite{BESIII:2020xfc}.

Analogous rescattering effects near other thresholds (e.g., $D\bar{D}^*$ and $D^*\bar{D}^*$) are expected due to similar interactions in the heavy-quark limit. The $\psi(4040)$, located close to the $D^*\bar{D}^*$ threshold ($\Delta m \approx 20~\mathrm{MeV}$) and having a large $D^*\bar{D}^*$ branching fraction, may also exhibit enhanced nonopen-charm hadron decays. Furthermore, similar threshold structures could appear in the hadronic decays of charmoniumlike structures such as $Z_c(3900)$~\cite{BESIII:2013ris,Belle:2013yex} and $Z_c(4020)$~\cite{BESIII:2013ouc}, where cusp effects may mimic resonant behavior~\cite{Chen:2011xk,Chen:2013wca,Chen:2013coa,Swanson:2014tra}.

Inclusion of the $h_c\pi\pi$ channel yields a resonancelike structure $R(3810)$ at $\sqrt{s} = 3.81~\mathrm{GeV}$ in the $e^+e^- \to h_c\pi\pi$ cross section, visible in nonopen-charm hadron but not open-charm channels, with minimal impact on the $\psi(3770)$ pole. The small fitted value $\Lambda_{h_c} = 0.04~\mathrm{GeV}$ implies the coupling $f_{\psi(1D),h_c\pi\pi}(\vect{q},\vect{p})$ is significant only at low momenta ($q, p \lesssim 0.04~\mathrm{GeV}$), consistent with soft-gluon emission in QCD multipole expansion~\cite{Kuang:2006me}. Further experimental study of $e^+e^- \to h_c\pi\pi$ is warranted

\noindent{{\it Summary}.---}The discovery of numerous charmonium and charmoniumlike states near open-charm thresholds has posed significant challenges to our understanding of nonperturbative quantum chromodynamics. In particular, the energy region around the $D\bar{D}$ threshold exhibits a proliferation of structures---$\psi(3770)$, $G(3900)$, $R(3760)$, $R(3780)$, and $R(3810)$---that defy conventional quark model classifications and suggest the presence of complex coupled-channel effects. The $\psi(3770)$, nominally assigned as the $1^3D_1$ charmonium state, shows several anomalies including a mass shift from potential model predictions, a large nonopen-charm decay fraction, and a distorted line shape. Recent high-precision measurements of $e^+e^- \to \mathrm{hadrons}$ cross sections further reveal multiple near-threshold enhancements, indicating that a unified description beyond conventional hadron or exotic state interpretations is required.

In this paper, we develop a coupled-channel framework incorporating both open-charm and nonopen-charm hadron channels to simultaneously describe all five near-threshold structures. The model includes bare $c\bar{c}$ states---$\psi(2S)$, $\psi(1D)$, and $\psi(3S)$---with masses and transition matrix elements constrained by phenomenological and QPC model inputs. By considering couplings to $D^{(*)}\bar{D}^{(*)}$ channels and introducing rescattering via an effective $s$-channel potential, we reproduce the asymmetric line shape of the $\psi(3770)$ and the prominent $G(3900)$ peak in the $D\bar{D}$ cross section. The $R(3780)$ is unambiguously identified with the $\psi(3770)$ resonance. The threshold enhancement $R(3760)$ is explained as a consequence of $D\bar{D} \to \mathrm{nOCH}$ rescattering, without requiring a direct $c\bar{c}$ coupling. The $R(3810)$ structure is attributed to the coupling between the $\psi(1D)$ state and the $h_c\pi\pi$ channel.

Our analysis provides a natural explanation for the large nonopen-charm hadron branching fraction of the $\psi(3770)$ through its proximity to the $D\bar{D}$ threshold, which enhances rescattering decay amplitudes. The model successfully describes both open-charm, nonopen-charm and inclusive hadronic cross sections with a minimal set of parameters. The results underscore the necessity of coupled-channel effects in interpreting near-threshold phenomena and mark a transition from a static, quenched quark model to a dynamic, unquenched spectroscopy. The framework presented here is general and can be extended to bottomonium systems, offering a predictive approach for analyzing future high-precision data from facilities such as Belle II.

\noindent{{\it Acknowledgements}.---}R.Q.Q. is supported by the National Natural Science Foundation of China under Grant No.~12447124. This work is also supported by the Nation Natural Science Foundation of China under Grants No.~12335001 and No.~12247101, the 111 Center under Grant No.~B20063, the Natural Science Foundation of Gansu Province (No.~22JR5RA389, No.~25JRRA799), the Talent Scientific Fund of Lanzhou University, the fundamental Research Funds for the Central Universities (No.~lzujbky-2023-stlt01), the project for top-notch innovative talents of Gansu province and Lanzhou City High-Level Talent Funding.

\bibliography{refs}

\clearpage

\setcounter{section}{0}
\setcounter{equation}{0}
\setcounter{figure}{0}
\setcounter{table}{0}
\setcounter{page}{1}
\makeatletter

\renewcommand{\thesection}{S\arabic{section}}
\renewcommand{\theequation}{S\arabic{equation}}
\renewcommand{\thefigure}{S\arabic{figure}}
\renewcommand{\thetable}{S\arabic{table}}

\clearpage
\onecolumngrid

\section{Supplemental Materials}

This supplemental material provides additional details on the QPC amplitudes and description of nOCH cross section.

\subsection{The QPC amplitude}
\begin{figure}[htbp]
  \centering
    \includegraphics[width = 18cm]{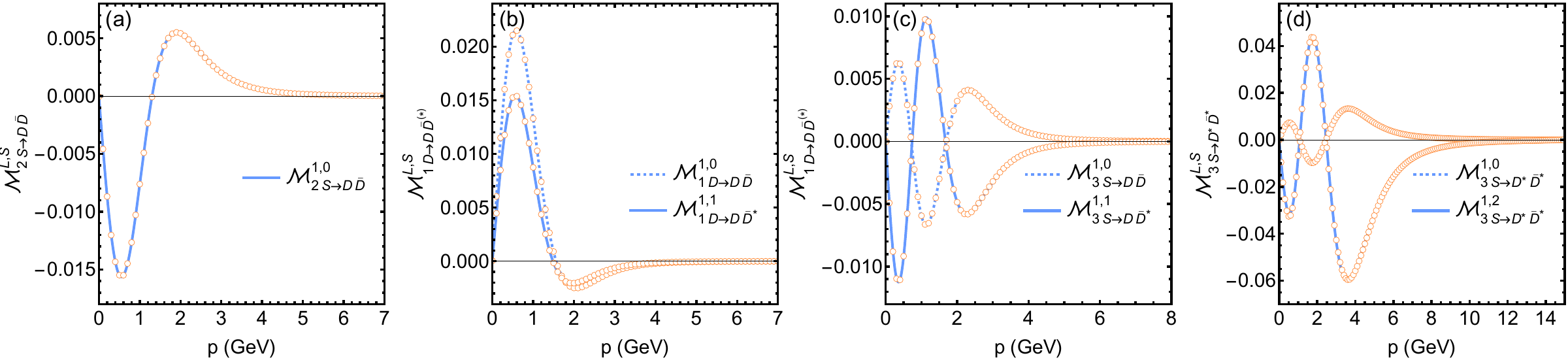}
    \caption{The calculated QPC partial wave amplitudes (orange circle points) for $\gamma = 1$ and fits to them in terms of analytical function (blue lines).}\label{fig:amp}
\end{figure}

The explicit form of various QPC amplitudes are calculated via the wave function of $c\bar{c}$ in the GI model.
In QPC model, the transition amplitude between charmonium $A$ and charmed meson pair $|BC;\vect{p}\rangle$ is described by 

\begin{eqnarray}
    \mathcal{M}^{QPC}_{M_{J_A},M_{J_B}M_{J_C}}(\vect{p}) &=& \gamma \sum_{M_{L_A}} \langle L_A M_{L_A}S_A M_{S_A}|J_A M_A\rangle\,  \langle L_{B}M_{L_B} S_B M_{S_B}|J_B M_B\rangle\, \langle L_C M_{L_C} S_C M_{S_C}|J_C M_C\rangle  \nonumber\\
    & & \times\, \langle 1m 1 -m|00\rangle\, \langle \chi_{S_B M_{S_B}}^{14}\chi_{S_C M_{S_C}}^{32}|\chi_{S_A M_{S_A}}^{12}\chi_{1-m}^{34} \rangle\, \langle \phi_B^{14}\phi_C^{32}|\phi_A^{12}\phi_0^{34}\rangle   \nonumber\\
    & & \times\, Ip(\vect{p}) \,,
\end{eqnarray}
where $\chi_i$ and $\phi_i$ are the spin and flavour wave functions of mesons. The $Ip(\vect{p})$ is a integral determined by the overlap of wave functions between initial and final states,
\begin{eqnarray}
    Ip(\vect{p}) = \int d^3q\, \psi^*_{B,L_B,M_{L_B}}\left( \frac{m_q}{m_c+m_q}\vect{p} + \vect{q} \right) \psi_{C,L_C,M_{L_C}}^*\left( \frac{m_q}{m_{\bar{c}}+m_q}\vect{p} + \vect{q} \right) \psi_{A,L_A,M_{L_A}}(\vect{p} + \vect{q})\mathcal{Y}_1^m(\vect{q}) \,,
\end{eqnarray}
where $\vect{p}$ is the CM momentum of $BC$ system, $m_c$ and $m_q$ are the quark masses. $\mathcal{Y}_1^m$ is the solid spherical harmonic function. It is convenient to express the amplitude in terms of partial wave amplitudes
\begin{eqnarray}
    \mathcal{M}^{QPC}_{M_A,M_B,M_C}(\vect{p}) = \sum_{L,S}\mathcal{M}_{A\to BC}^{L,S}(p)\langle L M_L;S M_S|1,1\rangle \langle S_1 M_1;S_2 M_2|S M_S\rangle Y_L^{M_L*}(\Omega) \,.
\end{eqnarray}
The calculated nonzero partial wave amplitudes when $\gamma=1$ are shown in Fig.~\ref{fig:amp}. For convenience, we use analytical functions of the form $\sum_i g_i p^L e^{-a_i p^2}$ to take the analytical approximation to these partial wave amplitudes, so the obtained $T$-matrix can be analytically continued to the second Riemann sheets. The fitted analytical function show perfect match with numerical results for real $p$, and it should give an good analytical continuation as long as the complex $p$ is not too far from the real axes.

Using the analytical $\mathcal{M}^{L,S}(p)$, the continuation of the integral $I_k(E)$ given in Eq.~\eqref{eq:Ik} is given by
\begin{eqnarray}
    I_k(E^{II}) = I_k(E^I) -i\,2\pi \mu_k p^{II} f_k(p^{II})f_k^\dagger(p^{II}) \,,
\end{eqnarray}
where $p^{II}=\sqrt{2\mu_k(E^{II}-m^{\mathrm{th}}_k)}$ is the momentum of $k$-th channel in the lower complex plane.

\subsection{The parameter dependence on $\Lambda_{nOCH}$ and $\Lambda_{h_c}$ }

\begin{figure}[htbp]
  \centering
    \includegraphics[width = 15cm]{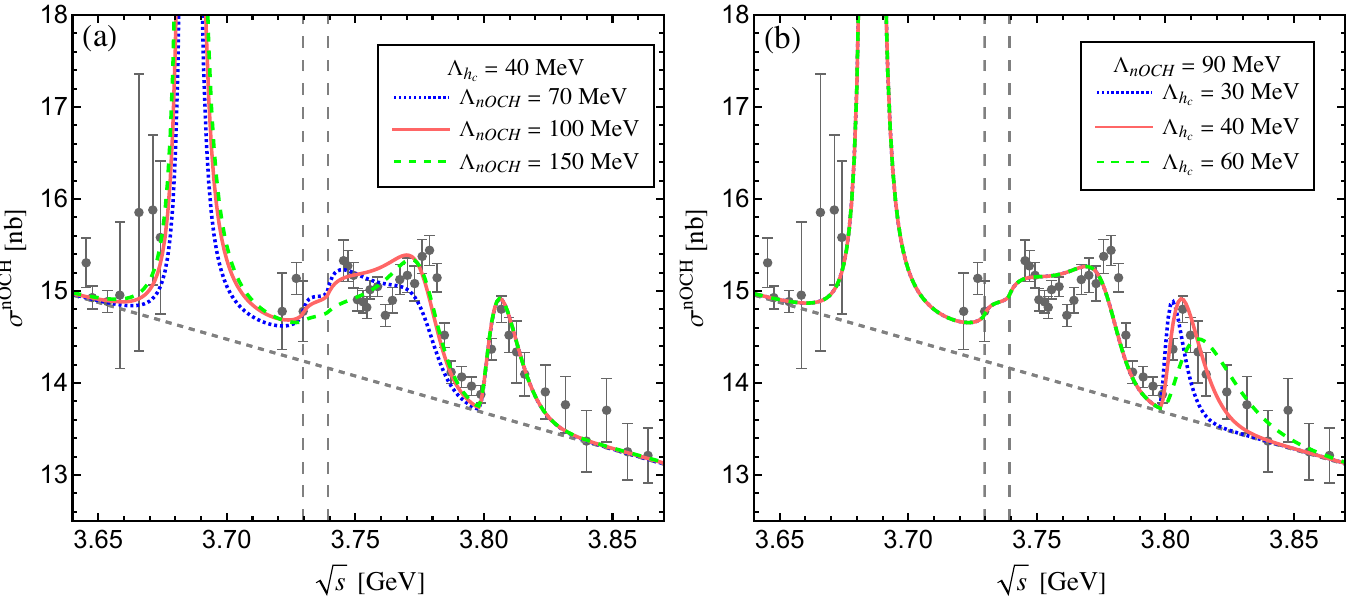}
    \caption{(a) The non-open-charm cross section for different $\Lambda_{nOCH}$, where $\Lambda_{h_c} = 40$ MeV. (b) The non-open-charm cross section for different $\Lambda_{h_c}$, where $\Lambda_{nOCH}=90$ MeV. }\label{fig:Lambda}
\end{figure}

The cross section of the non-open charm channels is induced by the $D\bar{D}\to nOCH$ and $\psi(1D)-h_c\pi\pi$ couplings, the main features of these coupling is their characteristic momentum scale, $\Lambda_{nOCH}$ and $\Lambda_{h_c}$, which is significant smaller than the momentum scale $\Lambda$ of the open charm coupling. Here, we quantify the uncertainty accociated with these mommentum scales. In Fig.~\ref{fig:Lambda}, we present the non-open-charm cross section for different $\Lambda_{nOCH}$ and $\Lambda_{h_c}$, where the corresponding coupling stregth $g_{nOCH}$ and $g_{h_c\pi\pi}$ are adjusted to reach an overall good description of data. From this analysis, we estimate the momentum scales to be
\begin{equation}
    70\; \mathrm{MeV} \lesssim \Lambda_{nOCH}\lesssim 150\; \mathrm{MeV}\,,\quad  30\; \mathrm{MeV} \lesssim \Lambda_{h_c\pi\pi}\lesssim 60\; \mathrm{MeV}\,.
\end{equation}
These scales reflect the underlying mechanism of the respective processes. The $\Lambda_{nOCH}$ set the momentum scale between $D\bar{D}$ in the $D\bar{D}\to nOCH$ process, it is suppressed unless $\Lambda_\text{nOCH}$ is small compared to the reduced mass of the $D\bar{D}$ system. Similarly, the $\Lambda_{h_c}$ reflect the momentum scale of $\psi(1D)-h_c\pi\pi$ coupling. Given that the bare mass of $\psi(1D)$ (3.82 GeV) is only about 100 MeV above the $h_c\pi\pi$ mass threshold, we expect only tens of MeV remaining momentum for the $h_c\pi\pi$ system.

\subsection{The complete LS equation for nOCH channel}

\begin{figure}[htbp]
  \centering
    \includegraphics[width = 15cm]{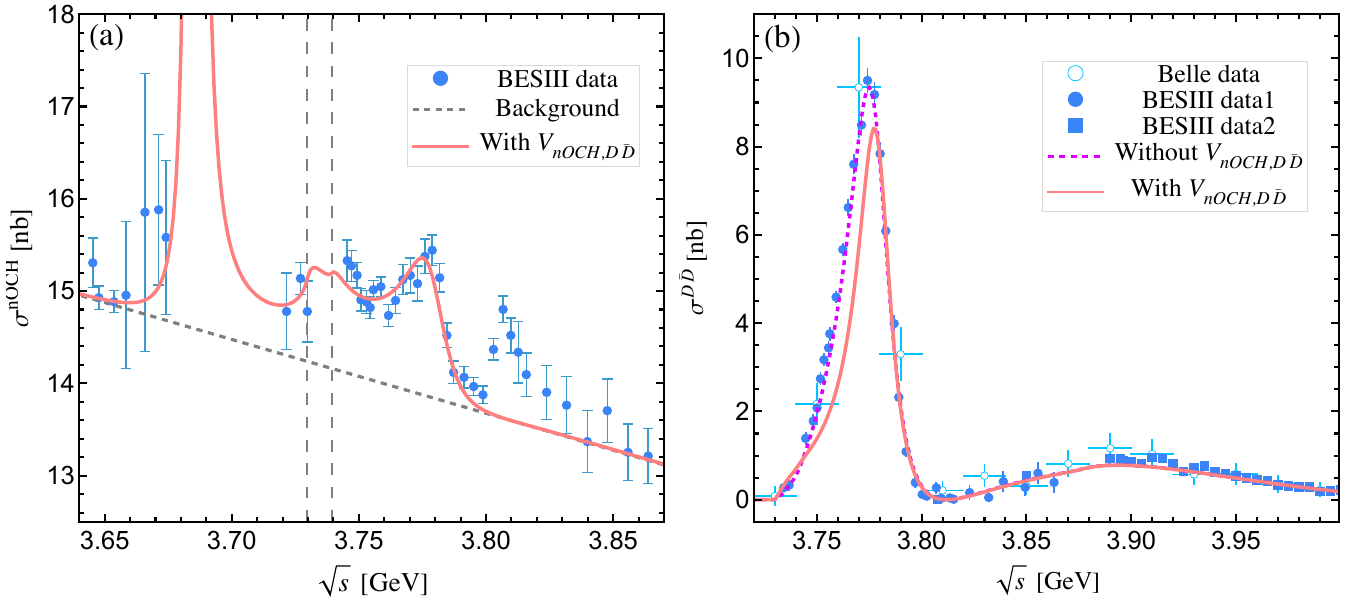}
    \caption{(a) The non-open-charm cross section calculated via LS equations when including the $V_{nOCH,D\bar{D}}$ potential. (b) The $D\bar{D}$ cross section calculated via the LS equation with/without $V_{nOCH,D\bar{D}}$ potential.}\label{fig:LS}
\end{figure}

To obtain the $T_{nOCH,e^+e^-}$, we approximate the $T_{D\bar{D},e^+e^-}$ using the amplitude in Eq.~\eqref{eq:TDD}, which is derived without the $V_{nOCH,D\bar{D}}$ interaction. Stirctly speaking, the $T_{D\bar{D},e^+e^-}$ should be solved via the complete set of LS equations
\begin{eqnarray}
    T_{D\bar{D},e^+e^-} &=& V_{D\bar{D},e^+e^-} + V_{D\bar{D},D\bar{D}}G^{D\bar{D}}T_{D\bar{D},e^+e^-} + V_{D\bar{D},D\bar{D}^*}G^{D\bar{D}^*}T_{D\bar{D}^*,e^+e^-} + V_{D\bar{D},D^*\bar{D}^*}G^{D^*\bar{D}^*}T_{D^*\bar{D}^*,e^+e^-} \nonumber\\
        &&  + V_{D\bar{D},nOCH}G^{nOCH}T_{nOCH,e^+e^-} \,,\\
    T_{D^{*}\bar{D}^{(*)},e^+e^-} &=& V_{D^{*}\bar{D}^{(*)},e^+e^-} + V_{D^{*}\bar{D}^{(*)},D\bar{D}}G^{D\bar{D}}T_{D\bar{D},e^+e^-} + V_{D^{*}\bar{D}^{(*)},D\bar{D}^*}G^{D\bar{D}^*}T_{D\bar{D}^*,e^+e^-} + V_{D^{*}\bar{D}^{(*)},D^*\bar{D}^*}G^{D^*\bar{D}^*}T_{D^*\bar{D}^*,e^+e^-} \\
    T_{nOCH,e^+e^-} &=&  V_{nOCH,D\bar{D}}G^{D\bar{D}}T_{D\bar{D},e^+e^-}\,.
\end{eqnarray}
By ignoring the $V_{nOCH,D\bar{D}}$, the solution to this system reduces to the form given in Eq.~\eqref{eq:TDD}. The complete LS equation can also be solved if the explicit momentum dependence of $V_{nOCH,D\bar{D}}(\vect{p}_{nOCH},\vect{p}_{D\bar{D}})$ is known. Assuming the form
\begin{equation}
    V_{nOCH,D\bar{D}}(\vect{p}_{nOCH},\vect{p}_{D\bar{D}}) = g_{nOCH,D\bar{D}} p_{nOCH}Y_1^{1}(\Omega_{nOCH})e^{-p_{nOCH}^2/\Lambda_1^2}p_{D\bar{D}}Y_1^{1*}(\Omega_{D\bar{D}})e^{-p_{D\bar{D}}^2/\Lambda_2^2} \,,
\end{equation}
we can solve the LS equation numerically. Here, we use the $g_{nOCH}=7.5\; \mathrm{GeV}^{-3/2}$, $\Lambda_1 = 1.58\; \mathrm{GeV}$ and $\Lambda_2 = 0.2\; \mathrm{GeV}$, while all other potentials parameters are same as those in Table~\ref{tab:parameter}. The result $e^+e^-\to D\bar{D}$ and $e^+e^-\to nOCH$ cross section is shown in Fig.~\ref{fig:LS}. The lineshape of non-open-charm cross section in the energy region 3.73-3.80 GeV is reproduced more accurately when the complete LS equation is considered. Furthermore, the $V_{nOCH,D\bar{D}}$ interaction have only small influence the resonance parameter of $\psi(3770)$.

\end{document}